\documentclass[a4paper,11pt]{article}

\usepackage{jcappub} 

\usepackage[T1]{fontenc} 
\graphicspath{{./fig/}}
\usepackage{bm}

\newcommand{\tildeH}{\tilde{H}}

\newcommand{\EQ}{\begin{equation}}
\newcommand{\EN}{\end{equation}}
\newcommand{\EQA}{\begin{eqnarray}}
\newcommand{\ENA}{\end{eqnarray}}

\newcommand{\Eq}[1]{equation~(\ref{#1})}
\newcommand{\Eqs}[2]{equations~(\ref{#1}) and~(\ref{#2})}

\newcommand{\App}[1]{Appendix~\ref{#1}}

\newcommand{\Sec}[1]{section~\ref{#1}}

\newcommand{\Fig}[1]{figure~\ref{#1}}

\newcommand{\Tab}[1]{table~\ref{#1}}

\newcommand{\bra}[1]{\langle #1\rangle}

\newcommand{\meanrho}{\overline{\rho}}

\DeclareMathAlphabet\mathbfcal{OMS}{cmsy}{b}{n}
{}
{}

{}
{}

{}
{}
{}
{}
{}
{}
{}
{}
{}
{}
{}
{}
{}
{}

{}

{}

{}
{}
{}

\newcommand{\tildeh}{\tilde{h}}

\newcommand{\kk}{\bm{k}}

\newcommand{\qq}{\bm{q}}
\newcommand{\xx}{\bm{x}}

\newcommand{\uu}{\bm{u}}

\newcommand{\TT}{{\bm{T}}}
\newcommand{\nab}{{\bm{\nabla}}}

\newcommand{\SSSS}{\mbox{\boldmath ${\sf S}$} {}}

\newcommand{\ii}{{\rm i}}

\newcommand{\DD}{{\rm D} {}}

\newcommand{\dd}{{\rm d} {}}

\def\Sp{\mbox{\rm Sp}}

\def\Rey{\mbox{\rm Re}}

\def\EEGW{{\cal E}_{\rm GW}}

\def\OmGW{{\Omega}_{\rm GW}}

\def\EK{E_{\rm K}}

\def\cs{c_{\rm s}}

\def\hrms{h_{\rm rms}}

\def\kp{k_{\rm p}}

\def\urms{u_{\rm rms}}

\def\divu2m{(\nab\cdot\uu)_{\rm rms}}

\def\etad{\eta_{\rm d}}
\def\etai{\eta_{\rm i}}
\def\etae{\eta}

\title{\boldmath Shallow relic gravitational wave spectrum with acoustic peak}

\author[a,b]{Ramkishor Sharma,}
\author[c]{Jani Dahl,}
\author[a,b,d,e]{Axel Brandenburg}
\author[c,f]{and Mark Hindmarsh}

\affiliation[a]{Nordita, KTH Royal Institute of Technology and Stockholm University,
Hannes Alfv\'ens v\"ag 12, 10691 Stockholm, Sweden}
\affiliation[b]{The Oskar Klein Centre, Department of Astronomy,
Stockholm University, 10691 Stockholm, Sweden}
\affiliation[c]{Department of Physics and Helsinki Institute of Physics, PL 64, FI-00014 University of Helsinki, Finland}
\affiliation[d]{McWilliams Center for Cosmology \& Department of Physics,
Carnegie Mellon University, Pittsburgh, PA 15213, USA}
\affiliation[e]{School of Natural Sciences and Medicine, Ilia State University,
3-5 Cholokashvili Avenue, 0194 Tbilisi, Georgia}
\affiliation[f]{Department of Physics and Astronomy, University of Sussex, Brighton BN1 9QH, United Kingdom}

\emailAdd{ramkishor.sharma@su.se}
\emailAdd{jani.dahl@helsinki.fi}
\emailAdd{brandenb@nordita.org}
\emailAdd{mark.hindmarsh@helsinki.fi}

\abstract{
We study the gravitational wave (GW) spectrum produced by acoustic waves in the early universe, 
such as would be produced by a first order phase transition, focusing on the low-frequency side of the peak. 
We confirm with numerical simulations the Sound Shell model prediction of a steep rise with wave number $k$ of $k^9$ 
to a peak whose magnitude grows at a rate $(H/k_\text{p})H$, where $H$ is the Hubble rate and $k_\text{p}$ the peak wave number, set by the 
peak wave number of the fluid velocity power spectrum. We also show that hitherto neglected terms give  
a shallower part with amplitude $(H/k_\text{p})^2$ 
in the range $H \lesssim k \lesssim k_\text{p}$, which in the limit of small $H/k$ rises as $k$.  
This linear rise has been seen in other modelling and also in direct numerical simulations.
The relative amplitude between the linearly rising part and the peak therefore depends on the peak wave number of the velocity spectrum and
the lifetime of the source, which in an expanding background is bounded above by the Hubble time $H^{-1}$. 
For slow phase transitions, which have the lowest peak wave number and the loudest signals, the acoustic GW peak appears as a localized enhancement of
the spectrum, with a rise to the peak less steep than $k^9$. 
The shape of the peak, absent in vortical turbulence, 
may help to lift degeneracies in phase transition parameter estimation at future GW observatories. 
}

\begin{document}
\maketitle
\flushbottom

\section{Introduction}
\label{sec:intro}

Relic gravitational waves (GWs) from the early universe can reveal
valuable information about the underlying generation mechanism and thus
the physical conditions at that time \cite{maggiore2000, chiara2020review,
NANOGrav2023}.
A particularly interesting epoch is the electroweak (EW) era, which may
have involved a first-order phase transition \cite{KIRZHNITS1976195,
coleman1977, linde1983, markreview2020}.
A first-order phase transition is characterized by the formation, expansion, and
subsequent merging of bubbles containing the low-temperature phase, leading to a
transition of the entire universe to a new phase, and the release of latent heat.
During this process, the kinetic energy transferred to the plasma can be a
considerable fraction of the total available energy and would therefore
be an important source of GWs, peaked at a frequency set by the inverse of the mean bubble spacing.
For GWs from the electroweak phase transition, the relevant frequencies
lie in the mHz range, which is accessible to the Laser Interferometer
Space Antenna (LISA) \cite{Audley:2017drz}.
Studies of GWs from that epoch have therefore attracted considerable attention.

The generation of GWs during a phase transition 
is often divided into three stages:
(i) the bubble collision phase, (ii) the acoustic 
phase, and (iii) the turbulence phase. 
The contribution from the bubble collision phase is typically
small compared to that from the other two stages, 
except in the case of a vacuum phase transition, where 
bubble collision becomes the only source of GW production
\cite{PhysRevD.45.4514,Kosowsky:1992vn,Huber:2008hg,Cutting:2018tjt,Lewicki:2020jiv}.
There has been extensive research on GW production by fluid flows in the early Universe 
through both numerical simulations
\cite{Giblin:2014qia,mark2014, 2015PhRvD..92l3009H, 2017PhRvD..96j3520H, RoperPol+20, cutting2020, Auclair:2022jod, Dahl:2021wyk, Jinno:2022mie} 
and semi-analytical or analytical models
\cite{Gogo+07, Caprini:2009yp, Caprini:2009fx, Hindmarsh_2016, Jinno:2017fby, Konstandin:2017sat, sigl2018, RoperPol:2022iel, SB22, Lewicki:2022pdb, Cai:2023guc}.
For phase transitions where the kinetic energy fraction is small, or for those proceeding by 
detonations, simulations indicate that the kinetic energy in 
vortical modes is subdominant compared to the
acoustic or compressional modes \cite{cutting2020}.
This paper concerns the acoustic contribution to the GW spectrum, and focuses specifically on its  
shape to the low-frequency side of the peak 
about which there is some uncertainty in the literature.

Relic GWs can be characterized by their normalized
energy spectrum, $\OmGW(k)$, where $k$ is the wave number, and
$\int\OmGW\,\dd\ln k$ is the 
fraction of the critical energy density of the universe in gravitational waves. 
The GW spectrum depends on the spectrum
of the hydrodynamic stress, which depends on the 
spectrum of the velocity field, how the stress is correlated in time, 
and on how quickly the stress appears \cite{Caprini:2009fx}.

Velocity spectra can be characterized by the wave number of the peak
$k_{\rm p}$ (also referred to as the energy-carrying wave number),
a declining inertial range spectrum for $k>k_{\rm p}$, and a rising 
subinertial range spectrum for $k<k_{\rm p}$. 
When the subinertial range spectrum is blue (steeper than that of white noise), the spectrum of
the stress is always white --- regardless of how blue the turbulence spectrum is 
\cite{BB20}.

Subject to certain assumptions about the 
time-dependence of the stress amplitude and its correlations, 
various arguments can be applied to conclude that the GW spectrum is peaked 
at wave numbers around $\kp$, and that 
the behavior of the GW spectrum on the low-frequency side of the peak should be linear \cite{Dufaux:2008dn,Caprini:2009fx, RoperPol+20,RoperPol:2022iel,Auclair:2022jod}.  The linear behavior applies 
down to a frequency set by the inverse source duration or the Hubble rate, whichever is the shorter.
Below this frequency the GW spectrum is white noise.

The most developed semi-analytical model for acoustic production of GWs is the Sound Shell model 
\cite{Hindmarsh_2016,Hindmarsh_2019}.
Studying fast transitions, where the peak wave number is much less than the inverse Hubble length, 
it was found that sound (or acoustic) waves produce a very steep $k^9$ GW spectrum at low frequencies.  
Other models, based on modelling the flow by expanding shells of shear stress, 
have indicated a $k^1$ spectrum at wave numbers below the peak. 
Direct numerical simulations of the sound wave phase of the phase transition
conducted in ref.~\cite{2017PhRvD..96j3520H,Jinno+23} show a peak, but no steep rise.
It remains unclear whether this discrepancy arises due to the 
inability of the simulations to capture the infrared part of the spectrum accurately, 
due to limited volume and duration, or for some other reason.

The purpose of the present work is therefore to reconsider the model
of ref.~\cite{Hindmarsh_2019} with a range of different approaches to
clarify the origin of the apparent conflict.
Recently, the authors of ref.~\cite{alberto} studied similar questions
and find a shallow part in the GW spectrum.
While their findings are consistent with ours, we provide a more detailed
investigation of how this shallow part appears and also complement our
findings by presenting direct numerical simulations of irrotational flows.
Furthermore, we compare the GW spectra generated by acoustic and vortical
flows in expanding backgrounds with similar stress spectra, confirming that the 
peak is still visible in the acoustic case, and distinguishes the two types of flow. 
It is worth noting that the acoustic spectra we study do not capture the
onset of turbulence, which may lead to new power laws emerging near the
peak \cite{Dahl:2021wyk}.

In \Sec{diff_approaches}, we discuss the different approaches used to
calculate the GW spectra originating from sound waves.
In \Sec{results}, we present our findings and compare the results
obtained using different approaches discussed in \Sec{diff_approaches}.
We discuss our findings in \Sec{conclusion}. 
In \App{OriginBackground} and \App{time_evolution_of_linear_part}, we give details on the origin of the
linearly rising GW spectrum, while in \App{TimeEvolutionKineticEnergy}
we report on a check on the effect of the growth rate of the stress.
In \App{appendixd}, we discuss an initial condition with non-uniform energy density.
Throughout this paper, we adopt units, 
where the speed of light is unity.

\section{Approaches to computing the gravitational wave spectrum}\label{diff_approaches}

\subsection{Gravitational wave spectrum}
\label{GravitationalWavesSpectrum}
Gravitational waves are represented by the transverse and traceless
components of metric perturbations.
In our analysis, we consider the background to be a homogeneous,
isotropic, and spatially flat expanding universe.
The metric describing such a universe with tensor perturbations can be
expressed as
\begin{equation}
ds^2 = a^2(\eta)\left[-d\eta^2 + (\delta_{ij} + h_{ij})dx^idx^j\right],
\end{equation}
where $a(\eta)$ represents the scale factor and $\eta$ denotes the conformal time.
The evolution of $h_{ij}$ is determined by the Einstein equation.
By employing this equation, we obtain the following linearized equation of motion for the Fourier space representation of $h_{ij}$,
\begin{equation}
\tilde{h}_{ij}'' + \frac{2a'}{a}\tilde{h}_{ij}' + k^2 \tilde{h}_{ij} = 16\pi G a^2 \tilde{\Pi}^{TT}_{ij}.
\end{equation}
Here, tildes symbolize quantities in Fourier space; this
convention is consistently used throughout this paper.
In the above equation, $a^2 \tilde{\Pi}^{TT}_{ij}$ represents the transverse
traceless part of the energy-momentum tensor $\tilde{\Pi}_{ij}$ of a GW source.
In terms of the stress tensor normalized by the energy density at the
initial epoch ($\rho_*$) and $\tildeH_{ij}=a \tildeh_{ij}$, the above equation
reduces to,
\begin{equation}
\tildeH''_{ij}+ \left(k^2-\frac{a''}{a}\right) \tildeH_{ij} =\frac{6 H_*^2 a_*^4}{a} \tilde{T}_{ij},
\end{equation}
where $\tilde{T}_{ij}=a^4 \tilde{\Pi}_{ij}/(a_*^4\rho_*)$ is the normalized stress
and $a_*$ represents the value of the scale factor at the initial epoch.
In the radiation-dominated era, using $a=a_*(\eta/\eta_*)$, and after
replacing $k \rightarrow k/(a_*H_*)$ and $\eta \rightarrow \eta/\eta_*$
($\eta_*$ and $H_*$ denote the conformal time and Hubble parameter at
the initial epoch, respectively), the above equation reduces to
\begin{equation}\label{hijeq}
\tildeH_{ij}''+k^2\tildeH_{ij} ={\cal G}(\eta) \tilde{T}_{ij}.
\end{equation}
Here, ${\cal G}(\eta)=6 a_*/\eta$.
In the limit of wave number large compared with the Hubble 
rate  ($k \to \infty$), 
one can make the approximation ${\cal G}(\eta) = 6 a_*/\eta_*$, equivalent to 
a non-expanding background. 
With our choice of units and scale factor,  ${\cal G}(\eta)=6$ in a non-expanding background 
(see ref.~\cite{RoperPol+20b} for details).

Usually, one is interested in estimating the GW spectrum at the present day, $\OmGW(k)$.
It represents the GW energy density per logarithmic wave number interval normalized
by the present-day critical energy density of the universe.
For the case when the source is active during the interval $\etai$ to
$\etae$, where $\etae$ denotes the time until the turbulence is active,
$\OmGW(k)$ is given by \cite{Chiara2009}
\begin{align}
\OmGW(k)&=\frac{3k^3}{ 4\pi^2}\left(\frac{a_*^2H_*}{a_0^2 H_0}\right)^2\int_{\etai}^{\etae} 
\int_{\etai}^{\etae} \frac{d\eta_1 d\eta_2}{\eta_1 \eta_2} \cos k(\eta_1-\eta_2)
U_T(k,\eta_1,\eta_2)\label{GWspec},
\end{align}
where $U_T(k,\eta_1,\eta_2)$ is the normalized fluid shear stress unequal
time correlator.
Here $H_0$ and $a_0$ denote the Hubble parameter and the scale factor at
the present epoch, respectively and $U_T(k,\eta_1,\eta_2)$
is defined through 
\begin{equation}
\langle \Tilde{T}_{ij}(\kk,\eta_1)
\tilde{T}^{ij}(\qq,\eta_2)\rangle=(2\pi)^3\delta(\kk-\qq)U_T(k,\eta_1,\eta_2).
\end{equation}

\subsection{The non-expanding universe case}

When the effective duration of the GW source is shorter than the Hubble
expansion time, it is possible to neglect the expansion of the universe
in estimating the GW energy spectrum.
In such cases, $\OmGW$ is given by
\begin{equation}\label{GWstatic}
\OmGW(k)=\frac{3k^3}{ 4\pi^2 }\left(\frac{a_*^2H_*}{a_0^2 H_0}\right)^2\int_{\etai}^{\etae} 
\int_{\etai}^{\etae} d\eta_1 d\eta_2 \cos k(\eta_1-\eta_2)U_T(k,\eta_1,\eta_2).
\end{equation}
The above expression is similar to the expression obtained in
ref.~\cite{Hindmarsh_2019} after substituting their equation (3.11)
into their equation (3.6).
The only difference is that in our case we have normalized the stress
tensor with the energy density of the initial epoch and the GW energy
density with the present-day critical energy density.

The main reason for considering the non-expanding case is that
we want to assess the validity of the approximations used in
ref.~\cite{Hindmarsh_2019}, where a non-expanding universe was assumed.
This allows us to provide a more detailed understanding of the origin
of particular features in the GW spectra.

\subsection{Calculation based on the sound shell model}
\label{Calculation}
The evaluation of the GW energy spectrum requires determining
$U_T(k,\eta_1,\eta_2)$
resulting from colliding sound waves generated by a phase
transition in the early universe.
The stress tensor for the fluid is given by
\begin{equation}
\Pi_{ij}=w \gamma^2 u_i u_j,
\end{equation}
where $w = \rho + p$ is the enthalpy density consisting of the
energy density $\rho$ and the pressure of the fluid $p$.
Here, $u_i$ and $\gamma=(1-\uu^2)^{-1/2}$ are the components of the
fluid 3-velocity and Lorentz factor, respectively.
The sound shell model of ref.~\cite{Hindmarsh_2019} assumes that the
expanding shells of pressurized plasma surrounding bubbles of the new phase 
continue to propagate after the phase transition is completed, each of
them acting as an initial condition for a sound wave.

The velocity field generated by a phase transition involving a large
number of expanding bubbles is then at each point a collection of sound
waves resulting from many of these sound shells.
The velocity can then be treated as a Gaussian random field.
Therefore, in calculating $U_T(k,\eta_1,\eta_2)$ for
a non-relativistic fluid using the standard method of expanding the
resulting connected four-point correlator using the Wick expansion,
any non-Gaussianities are thus ignored.
The four-point correlator then reduces to a sum containing products of
two-point velocity correlators.
The velocity correlator is assumed to be curl-free, like the flow around 
a single bubble, and to satisfy 
the linearized wave equation. 
In Fourier space, it then takes the form 
\begin{equation}
\bra{\tilde{u}_i(\qq,\eta_1) \tilde{u}^*_j(\qq',\eta_2)} = (2\pi)^3\delta(\qq-\qq')  \cos[\omega(\eta_1-\eta_2)]\,4\pi^2 q_i q_j \EK(q)/q^4,
\end{equation}
where (adopting conventions of non-relativistic fluid dynamics) $\EK(q)$ is the kinetic spectrum 
per {\em linear} wave number interval, $\omega=\cs q$ is the angular frequency,
and $\cs=1/\sqrt{3}$ is the sound speed in the plasma, assumed ultrarelativistic. 
The kinetic spectrum obeys $\int \EK dq = \bra{\uu^2}_V/2$, where $\bra{\,}_V$ represents a volume average.
We write the kinetic spectrum per logarithmic wave number explicitly as $q\EK(q)$. 

The result for $U_T(k,\eta_1,\eta_2)$ is equation~(3.34)
in ref.~\cite{Hindmarsh_2019}, which can be written as
\begin{equation}\label{stressinssm}
U_T(k,\eta_1,\eta_2)=\frac{16 \pi^2\bar{w}^2}{ k}
\int_0^{\infty} dq \int_{|q-k|}^{q+k} d\tilde{q}~\frac{q}{\tilde{q}^3} (1-\mu^2)^2 \EK(q) \EK(\tilde{q}) \cos[\omega(\eta_1-\eta_2)]  \cos[\tilde{\omega}(\eta_1-\eta_2)].
\end{equation}
where $\bar{w}=\bar{\rho}+\bar{p}$ is the mean enthalpy
density and $\mu=(q^2+k^2-\tilde{q}^2)/2kq$ is the cosine of the angle between $\kk$ and $\qq$. 
In our calculations, we assume a kinetic spectrum of the form
\begin{equation}
\EK(k) =\frac{3}{2\pi} \frac{\urms^2}\kp \frac{(k/\kp)^4}{1+(k/\kp)^6} \equiv \frac{3}{2\pi} \frac{\urms^2}{\kp}  \tilde{E}_{\rm K}(k),
\label{InitialSpectrum}
\end{equation}
with $\kp$ being the nominal peak wave number, $\urms$ the root-mean-square velocity, and $\tilde{E}_{\rm K}(k)$ the normalized kinetic spectrum.
Whenever we plot spectra from a simulation, we compute them for wave numbers sampled uniformly in $k$.
For the above form of the kinetic spectrum, the actual peak of the
spectrum is at $2^{1/6}\kp\approx1.12\,\kp$.
This form is consistent with irrotational and causal \cite{DC03} flows with shocks \cite{Dahl:2021wyk}, as appropriate for phase transitions. 
The peak wave number is inversely proportional to the mean bubble spacing.

We assume that the kinetic spectrum appears instantaneously, and remains at a constant amplitude. 
In \App{TimeEvolutionKineticEnergy}, we consider a simple model for the growth of kinetic energy during the phase transition.
By substituting $U_T(k,\eta_1,\eta_2)$ from \Eq{stressinssm}
into \Eq{GWstatic}, the GW spectrum can be obtained 
as
\begin{equation}\label{gwspeceqn}
\OmGW (k) 
= \Omega_0 \int_0^{\infty} dq \int_{|q-k|}^{q+k} d\tilde{q} \, \rho(k, q, \tilde{q})
\tilde{E}_{\rm K}(q) \tilde{E}_{\rm K}(\tilde{q}) \Delta(\eta, \etai, k, q, \tilde{q}),
\end{equation}
where
\begin{equation}
\rho(k, q, \tilde{q}) = \frac{[4k^2q^2 - (q^2+k^2 - \tilde{q}^2)^2]^2}{16 q^3 \tilde{q}^3 k^2} \end{equation}
and
\begin{equation}
    \Omega_0= \frac{54(1+c_s^2)^2}{\pi^2} \left( \frac{ u_\text{rms}^2}{\kp} \right)^2 \left( \frac{a_*^2 H_\star}{a_0^2 H_0} \right)^2=3.8~ \frac{g_*}{106.75} \frac{3.36}{g_0} \left(\frac{g_{0s}}{3.94}\frac{106.75}{g_{*s}}\right)^{4/3}\Omega_r \left( \frac{u_\text{rms}^2}{\kp} \right)^2.
\end{equation}
Here, $\Omega_r$ represents the radiation energy density fraction at
the present epoch, $g_{*s}$ and $g_{0s}$ denote the effective degrees
of freedom in entropy at the initial and the present epoch and $g_{*}$
and $g_{0}$ are the corresponding effective degree of freedom in energy
density.
In \Eq{gwspeceqn}, the time dependence is embedded into a kernel
$\Delta$, which now has the form
\begin{align}
\label{e:GWKerDef}
\Delta(\eta, \etai, k, q, \tilde{q}) &= \frac{1}{2}\int_{\etai}^\eta d\eta_1\int_{\etai}^\eta d\eta_2 \,
\sigma(\eta_1, \eta_2) \cos[k(\eta_1 - \eta_2)] \cos[c_s q(\eta_1 - \eta_2)]\cos[c_s \tilde{q}(\eta_1 - \eta_2)] \\
&= \frac{1}{8} \sum_{\pm\pm} \int_{\etai}^\eta d\eta_1\int_{\etai}^\eta d\eta_2 \, \sigma(\eta_1, \eta_2)
\cos[\omega_{\pm\pm}(\eta_1 - \eta_2)],
\end{align}
where $\omega_{\pm\pm} = k \pm c_s q \pm c_s \tilde{q}$, and the two different previously discussed cases regarding the expansion of the universe have been
taken into account by defining
\begin{equation}
\sigma(\eta_1, \eta_2) = 
\begin{cases}
    1 \quad \text{without expansion}, \\
    (\eta_1 \eta_2)^{-1} \quad \text{with expansion}.
\end{cases}
\end{equation}
The evaluation of the two time integrals can be carried out analytically in both cases.
For the non-expanding (i.e.~Minkowski) background, the result can be written in the form
\begin{equation}
\label{e:NonExpDel}
\Delta_\text{M} 
(\eta, \etai, k, q, \tilde{q})
= \frac{1}{2} \sum_{\pm\pm} \left(\frac{\sin[\omega_{\pm\pm} (\eta - \etai)/2]}{\omega_{\pm\pm}}\right)^2,
\end{equation}
while for the radiation-dominated expanding universe the result is
\begin{equation}
\label{e:ExpDel}
 \Delta_\text{exp} (\eta, \etai, k, q, \tilde{q}) = \frac{1}{8} \sum_{\pm\pm} \left\{ \left[\operatorname{Ci}{\left(\omega_{\pm\pm} \eta \right)} - \operatorname{Ci}{\left(\omega_{\pm\pm} \eta_{0} \right)}\right]^{2} + \left[\operatorname{Si}{\left(\omega_{\pm\pm} \eta \right)} - \operatorname{Si}{\left(\omega_{\pm\pm} \eta_{0} \right)}\right]^{2} \right\} \, ,
\end{equation}
where $\operatorname{Si}$ and $\operatorname{Ci}$ are the sine and cosine integrals.
Hence, to obtain the shape of the spectrum, only the $q$ and $\tilde{q}$ integrals
need to be evaluated numerically. All other numerical prefactors have been absorbed
into the constant $\Omega_0$.

In the analytical study of the GW spectrum in the non-expanding case in ref.~\cite{Hindmarsh_2019}, 
the term with $\omega_{--}$ in Eq.~(\ref{e:NonExpDel}) was shown to grow linearly with conformal time after the $q$ integral 
is performed, and 
was argued to dominate. This is consistent with the growth observed in Minkowski space simulations 
\cite{mark2014,2015PhRvD..92l3009H,2017PhRvD..96j3520H,Jinno:2022mie}. 
The linearly growing term was shown to behave as $k^{2n-1}$ for a logarithmic kinetic spectrum $k\EK(k) \sim k^{n}$. 
With $n = 5$ for $k \ll \kp$ and $n = -1$ for $k \gg \kp$, 
the linearly growing contribution to the GW spectrum has a steep $k^9$ spectrum at low wave number and a 
$k^{-3}$ spectrum at high wave number. In between there is a peak at a similar wave number to that of the peak in the 
kinetic spectrum, $\kp$. 
In view of its origin, we call it an acoustic peak.

Here, we consider all terms in \Eqs{e:NonExpDel}{e:ExpDel}.
The integrations were carried out using the {\tt Nintegrate} routine in Mathematica
with the adaptive Monte Carlo method, which has proven efficient for the case at
hand. The initial time has been chosen so that $\etai = 1$ and the upper limit in
the $q$ integral has been taken to be a large finite value of approximately $10\,\kp$.
Spectra produced by
the adaptive Monte Carlo method have been compared with those produced by quadrature
integration schemes, and the results are found to be in a good agreement with each
other with respect to accuracy.
Our numerical analysis demonstrates that, below the peak, there is indeed a
slope of $k^{9}$, although it does not extend to very low wave numbers.
Instead, there is a transition to a linear spectrum at a point that
depends on the product $\kp\eta$ in the non-expanding case, 
and on $\kp$ in the expanding case.\footnote{Recall that in our choice of units $k$ is the ratio of the wave number to the Hubble length at the initial time.}
We will provide a detailed discussion of these findings in \Sec{result1}.

\subsection{Simulations with a kinematic velocity field}
\label{KinematicVelocityField}

To compute the GW field numerically, we need to evaluate
the hydrodynamic stress in regular time intervals, $\delta\eta$.
In this section, we construct a three-dimensional, time-dependent, random
irrotational velocity field in Fourier space, $\tilde{\uu}(\kk,\eta)$, as
\begin{equation}
\tilde{\uu}(\kk,\eta)=\ii\kk\,\tilde{\phi}(\kk)\,\cos[\omega(k)\eta],
\label{uukin}
\end{equation}
where $\tilde{\phi}(\kk)$ is the Fourier transformed scalar potential
of the velocity, $\omega(k)=\cs k$ is the dispersion relation for sound
waves, $k=|\kk|$ is the wave number.
The formulation in \Eq{uukin} ensures that at $\eta=0$, the pressure
is uniform.
We construct $\tilde{\phi}(\kk)$ such that $k\EK(k)$ has a $k^5$
subinertial range for $k<k_{\rm p}$ and a $k^{-1}$ inertial range for
$k>k_{\rm p}$.
The potential $\tilde{\phi}(\kk)$ includes a phase factor
$\exp\ii\varphi(\kk)$ with random phases in the range
$-\pi<\tilde{\phi}(\kk)<\pi$.
The amplitude of $\tilde{\phi}(k)$ is chosen such that the resulting
kinetic spectrum has a broken power law (for details, see section~III\,B
of ref.~\cite{axel2017}).

The horizon wave number at the time of generation is $k=1$.
In our calculations, we sometimes include wave numbers below the
horizon wave number in order to see the subsequent evolution from
the $\OmGW\propto k^3$ spectrum at early times to a linearly increasing
one at later times for $k\eta>1$.

At each time step, we Fourier transform $\tilde{\uu}(\kk,\eta)$
into real space, $\uu(\xx,\eta)$, and compute $\Pi_{ij}(\xx,\eta)$.
From this, we compute the transverse tracefree (TT) projection
$\tilde{T}_{ij}^{\rm TT}(\kk,t)$ in Fourier space and express it in terms 
of plus and cross polarization modes, denoted with the subscript $A \in \{+,\times\}$. 
We then compute the strain field polarizations $\tilde{H}_A(\kk,\eta)$ by solving \cite{RoperPol+20}
\begin{equation}
\left(\frac{\dd^2}{\dd\eta^2}+k^2\right)\tilde{H}_{A}(\kk,\eta)=
{\cal G}\,\tilde{T}_{A}(\kk,\eta),
\end{equation}
where ${\cal G}=6$ in all our calculations without expansion, and
${\cal G}=6/\eta$ in our calculations with expansion.
For the simulations, we use the {\sc Pencil Code} \cite{JOSS}, which is
a massively parallel public domain code, where the relevant equations
have already been implemented \cite{RoperPol+20b}.
The computational domain is a cube of size $L^3$, where $k_1=2\pi/L$
is the lowest wave number.

Although $\tilde{\uu}(\kk,\eta)$ can be computed with high precision, regardless
of the choice of $\delta\eta$, we cannot choose the timestep too large, because
otherwise the accuracy of $\tilde{H}(\kk,\eta)$ will become poor. 
The error in the solution for $\tilde{H}(\kk,\eta)$ is $O(\delta\eta^2)$,
and the coefficient in the error is smaller than the third-order accurate
solution of the hydrodynamic equations, when those are computed; see
\Sec{Direct} below.

\subsection{Direct numerical simulations with the Navier Stokes equation}
\label{Direct}

The purpose of the kinematic velocity fields discussed in
\Sec{KinematicVelocityField} is to bridge the gap between actual
turbulence, which is nonlinear, and the linear random sound waves
considered in ref.~\cite{Hindmarsh_2019}.
To model turbulence more realistically, we also solve the compressible
Navier-Stokes equations directly, i.e.,
\begin{align}
\frac{\DD\uu}{\DD\eta}&=\frac{2}{\rho}\nab\cdot\left(\rho\nu\SSSS\right)
-\frac{1}{4}\nab\ln\rho+\frac{\uu}{3}\left(\nab\cdot\uu
+\uu\cdot\nab\ln\rho\right),
\label{dudt} \\
\frac{\partial\ln\rho}{\partial\eta}
&=-\frac{4}{3}\left(\nab\cdot\uu+\uu\cdot\nab\ln\rho\right),
\end{align}
where $\DD/\DD\eta\equiv\partial/\partial\eta+\uu\cdot\nab$ is the
advective derivative, ${\sf S}_{ij}=(\partial_i u_j+\partial_j u_i)/2
-\delta_{ij}\nab\cdot\uu/3$ are the components of the strain rate 
tensor, and $\nu$ is the viscosity.

We construct our initial condition in Fourier space by multiplying
a random vector field with a superposition of a vortical and an
irrotational contribution,
\begin{equation}
Q_{ij}=Q_0\,\left[
(1-p)(\delta_{ij}-\hat{k}_i\hat{k}_j)+p\hat{k}_i\hat{k}_j\right],
\end{equation}
where $0\leq p\leq1$ quantifies the irrotational fraction.
This tensor can also be written as
\begin{equation}
Q_{ij}=Q_0\,\left[(1-p)\delta_{ij}-(1-2p)\hat{k}_i\hat{k}_j\right].
\end{equation}
In this work, we consider the extreme cases $p=0$ and $p=1$
for vortical and irrotational flows, respectively.
In all cases, the initial density is usually chosen to be uniform and equal to $\meanrho$. 
In the irrotational case, however, this leads to a collection of
standing instead of traveling waves.
We therefore also consider non-uniform initial energy density distributions in
accordance with the continuity equation where the initial
Fourier-transformed logarithmic density is given by
$\widetilde{\ln\rho}(k) = \kk \cdot \uu/(c_s^2 k)$.
We provide a detailed illustration of this in \App{appendixd}, where
we compare with the case of a uniform initial energy density.

The relevant parameters for our simulations are the root-mean-square (rms) velocity, the Reynolds number,
$\Rey=\urms/\nu \kp$,
and the wave number $\kp$ of the spectral peak of the
initial velocity spectrum.
We recall that $k=1$ corresponds to the horizon wave number at the 
initial time, $\etai=1$.

\section{Results}\label{results}

\subsection{Spectral evolution with time}\label{result1}

\begin{figure*}\begin{center}
\includegraphics[scale=0.8]{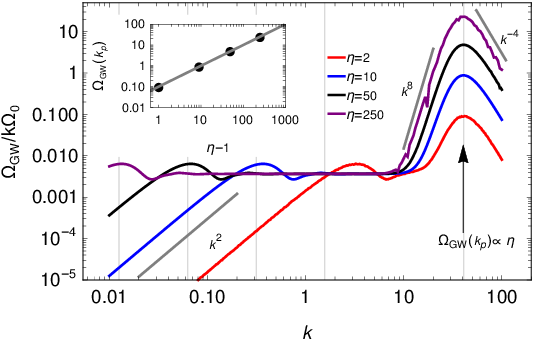}
\includegraphics[scale=0.57]{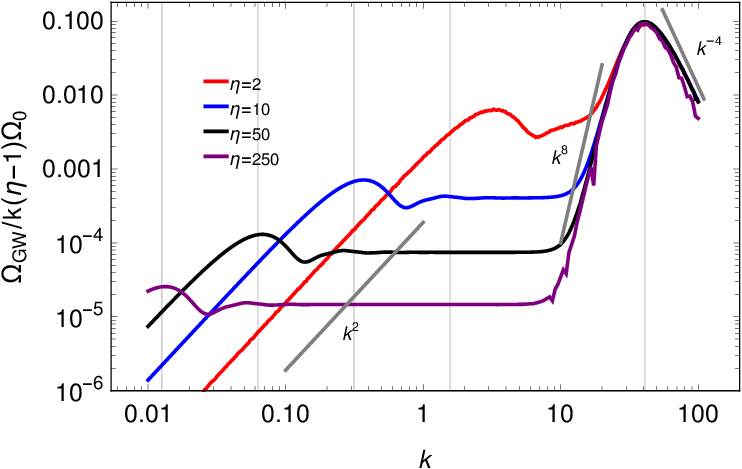}
\end{center}\caption[]{GW spectrum without expansion at different times.
The amplitude of the spectrum in the range $1<k<10$ is unchanged.
However, it continues increasing at $k=30$ with time $\propto \eta$.
By assuming that the GW spectra below the peak maintain the $k^8$
spectrum, the transition from $k^8$ to the flat regime moves toward
the left with a speed proportional to $\eta^{1/8}$.
The transition to $k^2$ propagates $\propto \eta$ in the horizontal
direction.
}\label{figure1}\end{figure*}

In \Fig{figure1}, we compare $\OmGW(k,\eta)/k\Omega_0$ for $\eta=2$,
10, 50, and $250$ in the non-expanding universe case. 
The division by wave number makes a linear growth with $k$ clear. 
The spectra were obtained by numerically integrating 
\Eq{gwspeceqn}.
In the calculations leading to \Fig{figure1}, we used \Eq{InitialSpectrum} for the
kinetic spectrum with $\kp=30$.
We see that $\OmGW(k,\eta)$ consists of a marked peak on top of a flat
part with $\OmGW(k)/k\propto k^0$.
As demonstrated below using numerical simulation, the peak does not appear for vortical flows.
The flat part corresponds to $\OmGW(k)\propto k$ and is therefore
also referred to as the shallow part of the GW spectrum.
The acoustic peak increases approximately linearly with time.
It does indeed have a $k^9$ slope below the peak over a short range
$10<k<30$.
Above the peak, $\OmGW(k,\eta)$ falls off approximately like $k^{-3}$.
This corresponds to $\OmGW(k,\eta)/k\propto k^{-4}$, as is indicated
in \Fig{figure1}.
This normalization shows more clearly the extent of the flat part
with $\OmGW(k,\eta)/k\propto k^{0}$.

\begin{figure*}\begin{center}
\includegraphics[scale=0.6]{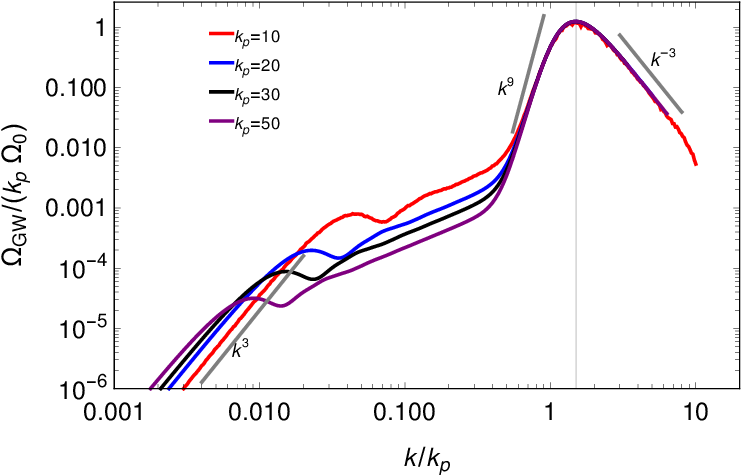}
\includegraphics[scale=0.6]{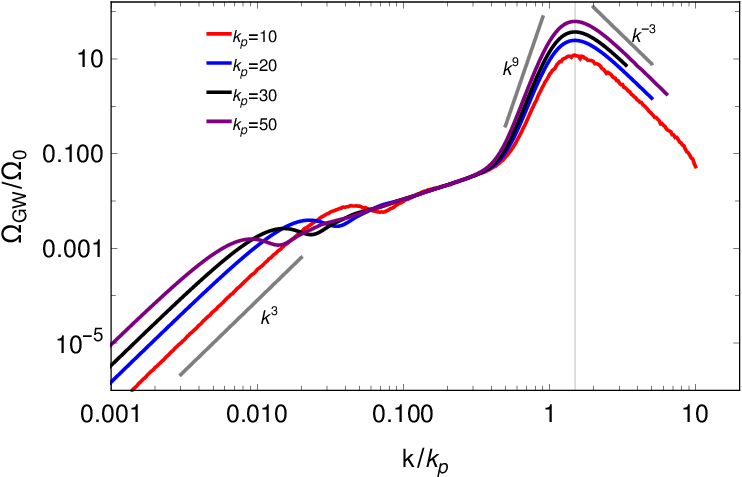}
\end{center}\caption[]{
Scaled GW spectrum for different values of $\kp$ at time $\eta=10$.
In the left panel, the spectrum is normalized by $\kp$ and by $\kp^2$
in the right panel.
}\label{GW_spec_peak_with_kp}\end{figure*}

The reason why we see the $k^9$ subrange only for a relatively short $k$
range is the presence of an approximately stationary and linearly rising
part with a shallow slope $\OmGW(k)\propto k$ for $1<k<10$.
This shallow rise was not present in the earlier work of
ref.~\cite{Hindmarsh_2019}.
As demonstrated in \App{OriginBackground}, the shallow part emerges
because of additional contributions to the time integral that are
not negligible in practice. 
Indeed, as mentioned in the introduction and discussed in  \App{OriginBackground}, 
a linearly increasing GW spectrum is expected on quite general grounds.
Furthermore, as time goes on, this linear GW spectrum
continues to extend to progressively lower $k$ values as the horizon
grows, smaller $k$ values become causally connected, and GWs begin to oscillate \cite{SB22}.

From \Fig{figure1}, it appears that the linear scaling part of
the GW spectrum remains unchanged with time.
This observation raises the question whether this constancy is
a result of the initial conditions we used.
However, it is important to note that the linear scaling part
develops gradually with the saturation occurring over a duration
equal to $2 \pi/k$ for the wave number $k$.
We have provided a time evolution of the $\OmGW/k \Omega_0$
for wave numbers within the region of linear scaling part in
\App{time_evolution_of_linear_part}.
In this analysis, we consider a constant kinetic spectrum, a choice
based on the assumption that the kinetic energy turns on within a time
duration shorter than $1/\kp$.

With time, the acoustic peak with its $\OmGW(k)\propto k^9$ rise continues to
grow, while the shallow part of the spectrum remains at constant amplitude. 
The wave number of intersection between the two power laws therefore 
moves toward lower wave numbers as $\eta^{1/8}$. 
The horizon wave number moves toward lower $k$ like $1/\eta$.
Therefore, the shallow part with $\OmGW(k)\propto k$ broadens with time.
However, since the height of the $\OmGW(k)\propto k^9$ feature continues
to grow, this steep part of the spectrum and the acoustic peak does indeed constitute an
important contribution 
to the spectrum.
To determine the ratio between the peak value of the spectrum and the
shallow part, it is necessary to estimate the duration of the interval
when the source remains active.
In this article, we primarily focus on the contribution from the
sound mode-dominated stage of the phase transition.
The sound mode contribution remains active until the development of
turbulence, and the relevant timescale for this is the eddy turnover
time corresponding to the peak of the spectrum.
This timescale is equal to the inverse product of the peak wave number of
the kinetic spectrum and the root-mean-square (rms) velocity of the fluid,
i.e., the nonlinear timescale $\eta_{\rm NL}=1/(\kp\urms)$.
If this timescale exceeds the Hubble expansion time,
the effect of expansion on GW production becomes important.

The Hubble expansion time 
sets an upper bound on
the effective duration
of the sound mode contribution.
We discuss this further in \Sec{result3}.
This scenario may be applicable to weak phase transitions and may be
of observational relevance.
For instance, when the peak of the kinetic spectrum occurs at a wave number
$10$ times greater than the Hubble rate and the rms fluid velocity
is 0.1, the value of $\eta_{\rm NL}$ relevant for turbulence development,
becomes comparable to the Hubble expansion time.
In \Sec{result3}, we provide a comparison considering the expansion for
this specific case.

In \Fig{GW_spec_peak_with_kp}, we show the compensated GW spectra $\kp \OmGW$
and $\kp^2 \OmGW$.
This allows us to see that the height of the peak of the GW spectrum
decreases like $1/\kp$ and the shallow part of the GW spectrum decreases
like $1/\kp^2$.
Thus, the height of the peak relative to the shallow part of the
spectrum grows linearly with $\kp$.

\subsection{Non-expanding case with synthesized random sound waves}\label{result2}

\begin{figure*}\begin{center}
\includegraphics[scale=1.2]{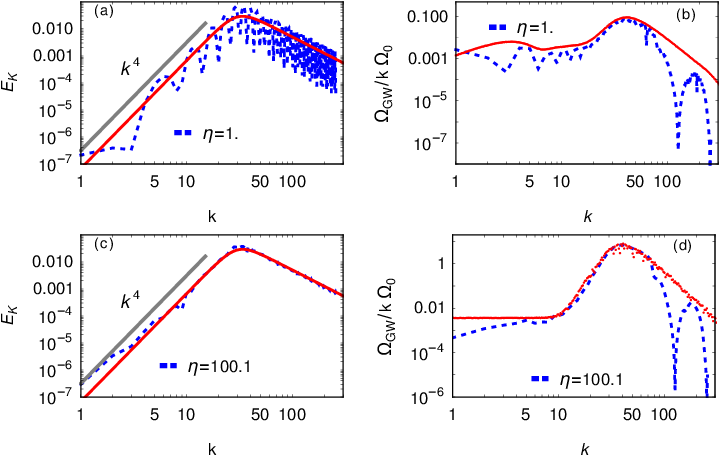}
\end{center}\caption[]{Kinetic spectra (left) and GW spectra (right), obtained numerically from synthesized sound waves in a non-expanding background, as described in \Sec{KinematicVelocityField}. 
Red curves show the semi-analytic spectra from the sound shell model, 
blue curves represent numerical spectra. 
Oscillations in the GW spectra at large $k$ are an artifact of large time steps.
}\label{Comparison}\end{figure*}

We now compare with the results from our model where the stress is
constructed from synthetic sound waves; see \Fig{Comparison}.
We should emphasize here that the GW energies from the semianalytical
method and the three-dimensional simulations agree with each other
rather well.
We see that the kinetic spectra have the expected $\EK\propto k^4$
subrange for $k<k_{\rm p}$, followed by a $k^{-2}$ subrange for $k>k_{\rm p}$.
(This $k^{-2}$ subrange is expected for acoustic turbulence, especially
in the presence of shocks \cite{Kadomtsev+Petviashvili73}, although in
the present case, shocks are only present when we solve the Navier-Stokes
equation.)

In \Fig{Comparison}, the GW spectra per linear wave number interval,
$\OmGW/k$, are shown at two different times $\eta=1.0$ and $\eta=100$.
At the early time $\eta=1$, the spectrum starts having a shallow subrange
and a steep part and these features become more pronounced at late times;
see \Fig{Comparison}(d).
It is evident from this figure that the steep and large wave number
part of $\OmGW$ is very well described by the spectra obtained by our
semianalytical model described in \Sec{Calculation}.

\subsection{Results from direct numerical simulations}

\begin{figure*}\begin{center}
\includegraphics[scale=1.25]{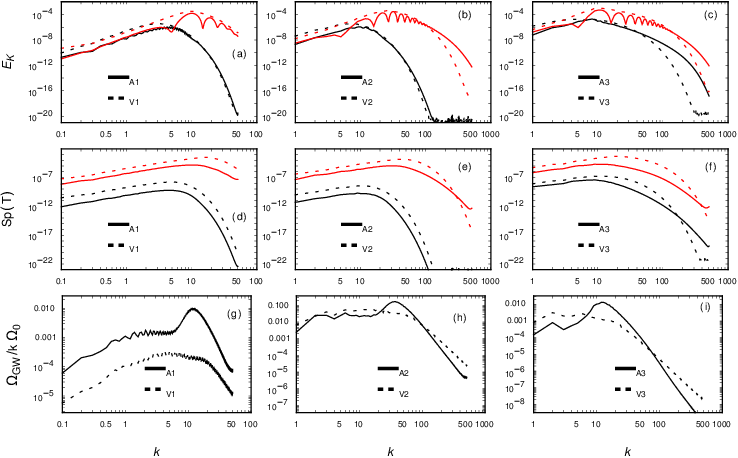}
\end{center}\caption[]{
Kinetic, stress and mean GW spectra for irrotational runs (solid lines) and vortical runs (dashed lines).
The red and black curves 
show spectra at $\eta=1.5$ and $\eta=41.0$, respectively.
Note the presence of the bump in all irrotational runs, 
absent in all vortical runs.
}\label{figure3}\end{figure*}

In \Fig{figure3}, 
we show the kinetic spectrum $\EK$, the stress spectrum 
$\Sp(\TT)\equiv U_T(k,\eta,\eta)\,k^2 /2\pi^2$,
and the compensated GW spectrum $\OmGW(k)/k \Omega_0$,
at different times for flows starting both from vortically and
irrotationally initialized velocity fields.
The parameters of these runs are summarized in \Tab{Tsummary}, 
where $\EEGW = \int\OmGW(k, \eta_*)/k\,\dd k$
is the total GW energy in units of the
critical density of the universe at the initial epoch.
The value of the Reynolds number $\approx40$ is small compared with what is expected
in reality. This is not a major concern 
as we are not interested in the development of turbulence, but in the relation between the GW spectrum and 
the kinetic spectrum.

Looking at \Fig{figure3}, we see that 
$\OmGW(k)/k \Omega_0$ 
has a bump at
$k\approx10...20$ for irrotational turbulence, while for vortical
turbulence there is no such bump.
Also, $\Sp(\TT)$ never shows a bump.
The bump in $\OmGW(k)$ is potentially an important characteristic of
acoustic turbulence in the GW spectra.
We also see that the inclusion of low wave numbers in the simulations
($k<1$ for Runs~A1 and V1) results in a clearer representation of the
$\OmGW(k)\propto k^3$ range, which is not visible for the other runs
where $k_1=1$ or larger.

The existence of a shallow part and the transition to $\OmGW(k)\propto k^3$ for
very small $k$ are independent of whether the turbulence is acoustic or vortical.
In all these cases, the height of the bump is about one order of magnitude.
The relative height is also independent of the overall amplitude of the turbulence.

At early times, some of the spectra show a wavy structure in $\EK(k)$.
These are not caused by numerical artifacts, but are a consequence
of having initialized a velocity pattern with zero energy density fluctuation, so 
all modes have the same phase.
This causes a modulation of the spectrum of the form $\cos k\xi(t)$,
where $\xi(t)=\cs t$ is the distance a sound wave has propagated in
the time $t$ since the initial condition was applied.
This type of wavy feature or ringing in the spectrum was explored and explained in
a different context in more detail in ref.~\cite{BN22}.
As shown in \App{appendixd}, however, this ringing phenomenon is strongly
reduced when the density is initialized such that the continuity equation
is obeyed.

\begin{table}[htb]\caption{
Summary of the parameters of the direct numerical simulations 
discussed in the paper.
All runs have a numerical resolution of $1024^3$ mesh points.
}\vspace{12pt}\centerline{\begin{tabular}{lcccccccc}
Run & $p$ & $A$ & $k_1$ & $\kp$ & $\nu$ & $\urms^2/2$ & $\EEGW$ & $\hrms$ \\
\hline
A1 &$1$&0.1 &$0.1$&$10$&$5\times10^{-3}$&$3.8\times10^{-2}$&$2.7\times10^{-5}$&$2.4\times10^{-3}$\\
V1 &$0$&0.1 &$0.1$&$10$&$5\times10^{-3}$&$7.5\times10^{-2}$&$1.1\times10^{-4}$&$6.9\times10^{-3}$\\
A2&$1$&$0.1$&$1.0$&$30$&$1\times10^{-3}$&$4.5\times10^{-2}$&$3.7\times10^{-6}$&$2.5\times10^{-4}$\\
V2&$0$&$0.1$&$1.0$&$30$&$1\times10^{-3}$&$9.1\times10^{-2}$&$1.4\times10^{-5}$&$8.1\times10^{-4}$\\
A3 &$1$&0.05&$0.1$&$10$&$5\times10^{-4}$&$1.2\times10^{-2}$&$7.7\times10^{-6}$&$6.5\times10^{-4}$\\
V3 &$0$&0.05&$0.1$&$10$&$5\times10^{-4}$&$2.4\times10^{-2}$&$8.4\times10^{-6}$&$2.0\times10^{-3}$\\
\label{Tsummary}\end{tabular}}\end{table}

\subsection{Effect of expansion in the sound shell model}\label{result3}

\begin{figure*}\begin{center}
\includegraphics[scale=0.6]{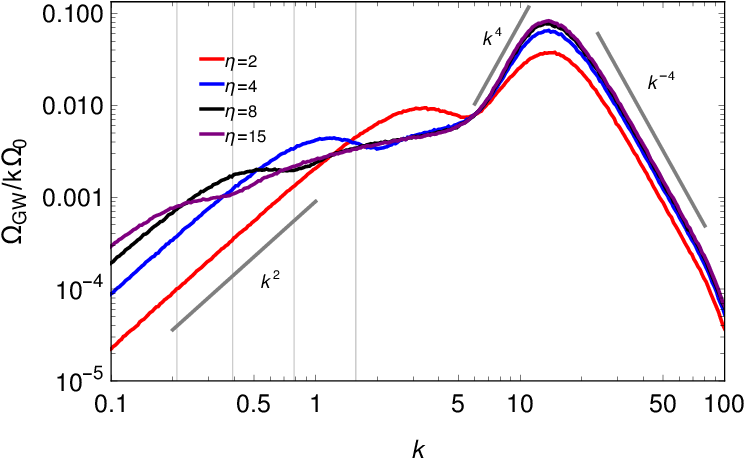}
\includegraphics[scale=0.6]{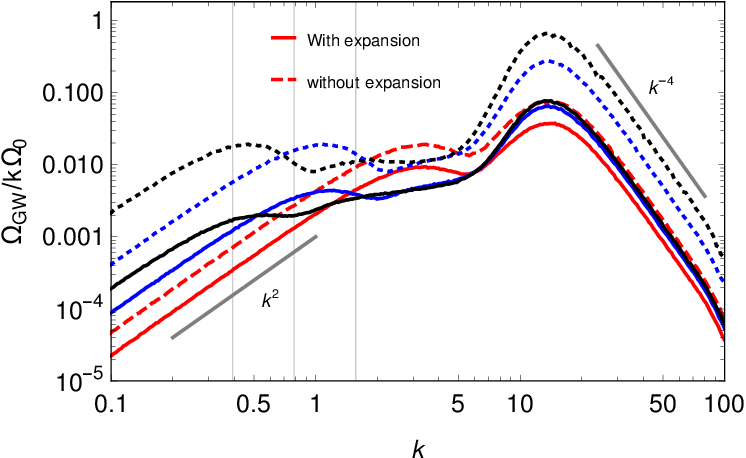}
\end{center}\caption[]{
Left: Compensated GW spectrum $\OmGW/k \Omega_0$ with expansion at different times obtained from the analytic model.
Note that the height of the peak over the shallow part is only about one
order of magnitude in $\OmGW/k$.
The gray lines indicate various slopes $\propto k^4$ and $k^{-4}$ for orientation
but are not meant to refer to any particular theoretical prediction. Right: Comparison of the GW spectra obtained for non-expanding (dotted curves) and expanding cases (solid curves).
}\label{figure4}\end{figure*}

As discussed in \Sec{GravitationalWavesSpectrum}, when the expansion of the
universe is taken into account, the effective duration of the acoustic source 
is limited and the steep part of the GW spectrum is much smaller.
This is shown in \Fig{figure4}, where we show $\OmGW(k)/k$ for $\kp=10$ at $\eta=5$,
15, and 30.
We see that the spectra do not change much at late times after $\eta=4$,
and that the results for $\eta=8$ and 15 are almost indistinguishable and close to those for $\eta=4$.
Furthermore, the height of the bump is limited to about one order of
magnitude.

\begin{figure*}\begin{center}
\includegraphics[scale=0.6]{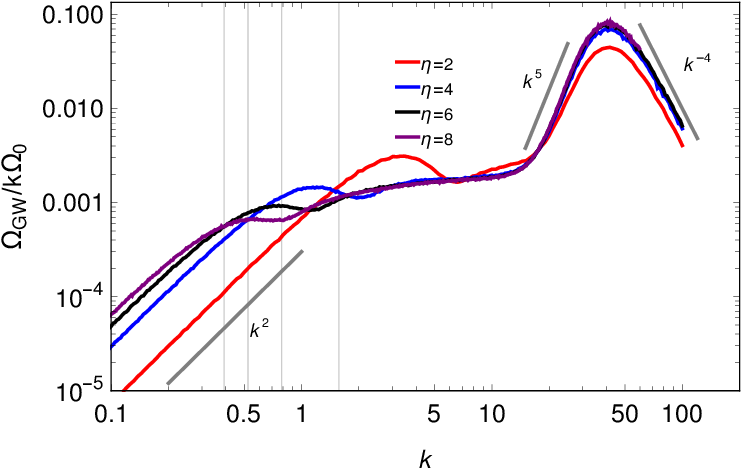}
\includegraphics[scale=0.6]{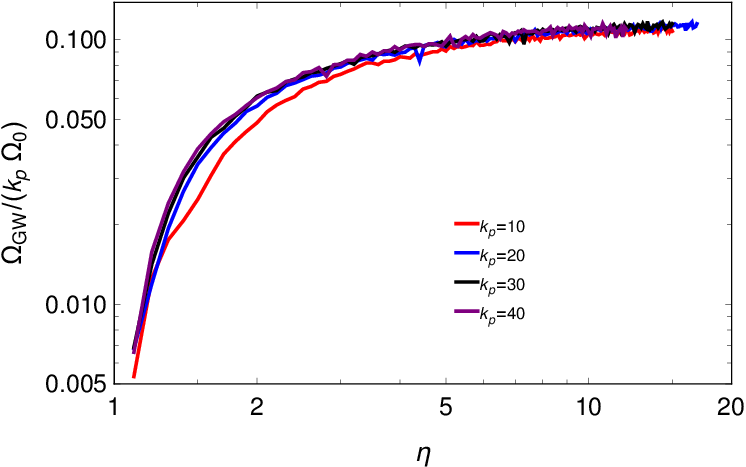}
\end{center}\caption[]{
Left: GW spectra for $\kp=30$ at times $\eta=2$, 4, 6, and 8 for cases with expansion included.
Right: time evolution of $\Omega(\kp)/(\kp \Omega_0)$
for the same case, but for different values of $\kp$.
}\label{spectrum_with_exp_kp30}\end{figure*}

\begin{table}[htb]\caption{
The spectral index of GW spectrum below the peak for the expanding case, computed in the sound shell model.}
\vspace{12pt}\centerline{\begin{tabular}{rcc}
      & \multicolumn{2}{c}{spectral index} \\
$\kp$ & $\etae=5$ & $\etae=10$ \\
\hline
  $10$ & $4.8$ & 5.0 \\
  $20$ & $5.6$ & 5.8 \\
  $30$ & $6.1$ & 6.3 \\
  $50$ & $6.6$ & 6.8 \\
  $70$ & $6.9$ & 7.0 \\
 $100$ & $7.2$ & 7.3 \\
 $200$ & $7.6$ & 7.7 \\
 $500$ & $8.0$ & 8.1 \\
$1000$ & $8.3$ & 8.3 \\
$2000$ & $8.4$ & 8.5 \\
\label{slope_table}\end{tabular}}\end{table}

The presence of a linearly rising $\OmGW(k)$ is not immediately apparent in \Fig{figure4}.
However, it becomes more pronounced in \Fig{spectrum_with_exp_kp30}
as we increase the value of $\kp$ to $30$.
We also estimate the spectral index of $\OmGW(k)$ just below its peak.
For this purpose, we fit a single power law within the 
$k$ values approximately in the range $0.5\, \kp$ to $0.9 \, \kp$.
The index is determined through least-squares fitting.
The results for the time $\etae=10$ are summarized in \Tab{slope_table}.
At time $\etae=10$, the spectrum is almost in a saturated state.
From \Tab{slope_table}, we conclude that, as we increase
the value of $\kp$, the GW spectrum below its peak tends toward a $k^9$ scaling.

In the right panel of \Fig{spectrum_with_exp_kp30}, we show the time evolution
of $\kp \Omega(\kp)$ for different values of $\kp$.
It is evident from the figure that all of these cases have a similar time
evolution and show saturation after $\etae \approx 5$, independently of
the value of $\kp$.

\section{Conclusions}\label{conclusion}

The present work has confirmed that there is indeed the very steep
$\OmGW\propto k^9$ rise to the peak in the GW spectrum for acoustic gravitational wave production,
as originally proposed in ref.~\cite{Hindmarsh_2019}.
However, it may not be very prominent in practice, because it is
superimposed on a stationary linearly rising part for all subhorizon
wave numbers above $k>1/\eta$ and below the spectral peak at $k\sim k_{\rm p}$.
The absence of this shallow part in earlier analytical calculations is a
consequence of having considered only the leading term in Eq.~(\ref{e:GWKerDef}),
as discussed in more detail in \App{ContributionFromSpm}. 
The height of the acoustic peak grows linearly in time, 
so it becomes distinct only at late times, and the $k^9$ subrange exists only in the range $\kp\eta^{-1/8} \lesssim k \lesssim \kp$. The $k^9$ subrange therefore becomes prominent only when the flow is long-lived and $\kp \gg 1$.

In \App{TimeEvolutionKineticEnergy} we considered the effect of the growth
rate of the hydrodynamic stress on the shape of the GW spectrum, using a
simple model where the stress approaches its final value exponentially
with time.
This allows analytic expressions for the time integrals to be maintained.
We find that the amplitude of the shallow part of the spectrum is reduced,
and decreases the slope below linear.
The steep rise to the peak is unaffected.

Given that the $k^9$ subrange takes time to emerge, we can ask whether indications of it have still
been seen in earlier work on acoustically generated GWs.
Acoustic GW production has been considered in several direct numerical simulations \cite{Hindmarsh_2019,RoperPol+20,BGKMRPS21,Jinno:2022mie}, 
but none has reported a $k^9$ slope.  However, a steepening 
of the slope leading to a peak with time has been observed in ref.~\cite{Jinno:2022mie}, up to approximately $k^5$. Our analysis indicates that this steepening would continue in a longer simulation with more dynamic range between the peak wave number of the velocity field and the inverse grid size.
On the other hand, vortical flows show no such peak. 
\cite{RoperPol+20,BGKMRPS21,Auclair:2022jod}. 
Both the presence and the shape of the peak therefore probe 
the nature of the flow (irrotational or vortical) and provide new information about the value of $\kp$, which may help lift potential degeneracies in parameter estimation identified in ref.~\cite{Gowling:2021gcy}. 
More comparative studies are needed before the shape of the peak in the GW spectrum can be used as a diagnostic tool in future observational studies with LISA.

\paragraph{Data availability.}
The source code used for the numerical solutions of this study, the
{\sc Pencil Code}, along with the additions included for the present
study, is freely available~\cite{JOSS}; see also ref.~\cite{DATA}
for the numerical data.

\acknowledgments
We thank David Weir for helpful discussions. 
Support through grant 2019-04234 from the Swedish Research Council
(Vetenskapsr{\aa}det), 
ERC HERO-810451 grant from the European Research Council,
NASA ATP grant 80NSSC22K0825, the Magnus Ehrnrooth foundation, and the Academy of Finland
is gratefully acknowledged.
We thank the Swedish National Allocations Committee for providing computing resources at the Center for Parallel
Computers at the Royal Institute of Technology in Stockholm.
Nordita is sponsored by Nordforsk.
We thank the authors of ref.~\cite{alberto} for sharing a draft of their paper, which 
addresses similar questions.

\appendix
\section{Origin of the shallow GW spectrum}
\label{OriginBackground}

\begin{figure*}\begin{center}
\includegraphics[scale=.9]{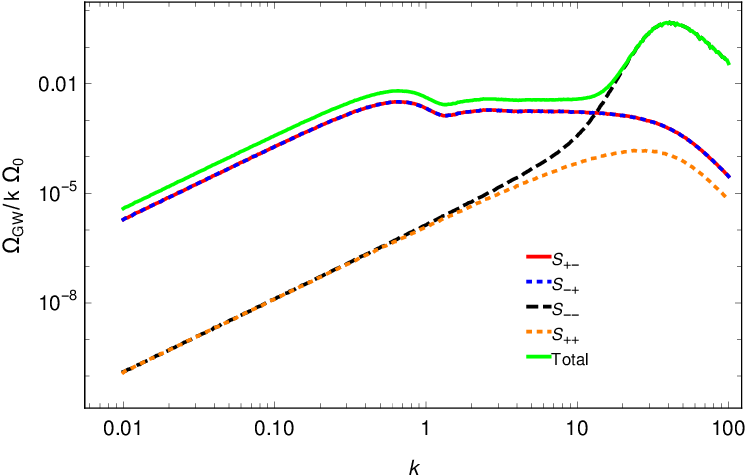}
\end{center}\caption[]{
Contributions to $\OmGW(k) h_0^2/k$ from the four terms in \Eq{FourTerms}
for $\etae=5$.
Note that the steep $\OmGW(k)\propto k^9$ contribution comes from the
$S_{--}$ term, while the linear scaling results from the $S_{+-}$ and
$S_{-+}$ terms.
}\label{contributions}\end{figure*}

Here, we elucidate the origin of the 
linearly varying GW spectrum at wave numbers below the peak. This linear behavior applies to wave numbers satisfying the condition $k \eta>1$, which is always below the peak.
As elucidated in \Sec{Calculation}, the GW spectrum is given by,
\begin{equation}\label{GWstatic_appendix}
\OmGW(k)=\Omega_0\int_0^{\etae} 
\int_0^{\infty} dq \int_{|q-k|}^{q+k} d\tilde{q}~\frac{q}{\tilde{q}^3} \rho(k, q, \tilde{q}) \tilde{E}_{\rm K}(q) \tilde{E}_{\rm K}(\tilde{q}) \Delta(\eta,\etai,k,\omega,\tilde{\omega}),
\end{equation}
where
\begin{equation}
 \Delta(\eta, \etai, k,\omega,\tilde{\omega})= \frac{1}{2}(S_{+-}^2+S_{-+}^2+S_{--}^2+S_{++}^2)
\label{FourTerms}
\end{equation}
with 
\begin{equation}
\label{e:Sdef}
    S_{\pm\pm}=\frac{\sin [(k\pm \omega\pm\tilde{\omega})(\etae-\etai)/2)]}{(k\pm\omega\pm\tilde{\omega})},
\end{equation}
in the non-expanding case, and in the $k\etai \to \infty$ limit of the expanding case.
This expression is the same as that of ref.~\cite{Hindmarsh_2019} in which there are eight terms, because
each of the four terms in \Eq{FourTerms} contributes twice owing to
a corresponding term with opposite sign in front of $k$.

The term relevant for the steep $k^9$ spectrum is $S_{--}^2$, which was argued to be 
the dominant term in ref.~\cite{Hindmarsh_2019}, as it produces a contribution linearly growing with conformal time $\eta$ (see \Fig{contributions}).
However, there are still contributions from $S_{+-}^2$ and $S_{-+}^2$,
which result equally in the formation of a shallow spectrum.
These terms were neglected in ref. \cite{Hindmarsh_2019}.

\label{ContributionFromSpm}

To see this, we estimate the GW spectrum in the range $k\etae\gg1$ for
wave numbers well below the peak $k/\kp \ll 1$. Hence $[k+c_s(q-\tilde{q})]\etae/2$ has a large
value and we can approximate $\sin^2[(k+c_s(q-\tilde{q}))\etae/2]$ by its average value $1/2$.
Using this, and taking the contribution only from the $S_{+-}$ term,
\Eq{GWstatic_appendix} reduces to,
\begin{equation}\label{eqnc1}
\OmGW^{+-}(k)=\Omega_0 
\int_0^{\infty} dq \int_{|q-k|}^{q+k} d\tilde{q} \rho(k, q, \tilde{q}) \tilde{E}_{\rm K}(q) \tilde{E}_{\rm K}(\tilde{q}) \frac{1}{4[k+c_s(q-\tilde{q})]^2}.
\end{equation}
For further analysis, we approximate the kinetic spectrum such that
it takes on zero values above the peak wave number $\kp$,
\begin{equation}\label{new_kinspec}
    \tilde{E}_{\rm K}(k)= \left\{\begin{array}{ll}
     (k/\kp)^4 , \quad & k\le \kp, \\
    0, & k > \kp.
    \end{array}\right.
\end{equation}
Then we define variables $x = q/k$, $y = \tilde{q/k}$, in terms of which the integral can be written as
\begin{equation}\label{eqnc1rescale}
\OmGW^{+-}(k)=\Omega_0 (k/k_\text{p})^8 
\int_0^{k_\text{p}/k} dx \int_{|x-1|}^{x+1} d y \rho(1, x, y) \frac{x^4 y^4}{4[1+c_s(x-y)]^2}.
\end{equation}
Considering the limit of large $\kp/k$, the integral is dominated by the upper limit on $x$, while the $y$ can be replaced by $x$, and $x-y$ is O(1). Hence, given that $\rho(1,x,x) \sim x^{-2}$, we have 
\begin{equation}\label{eqnc1final}
\OmGW^{+-}(k) \propto \Omega_0 (k/k_\text{p})^8 
\int_0^{k_\text{p}/k} dx x^6 \propto \Omega_0 ( k/\kp) .
\end{equation}
This linear scaling with $k/\kp$ is shown in \Fig{GW_spec_peak_with_kp}.

\section{Early time dependence of the shallow part of the GW spectrum}\label{time_evolution_of_linear_part}
\begin{figure*}\begin{center}
\includegraphics[scale=0.6]{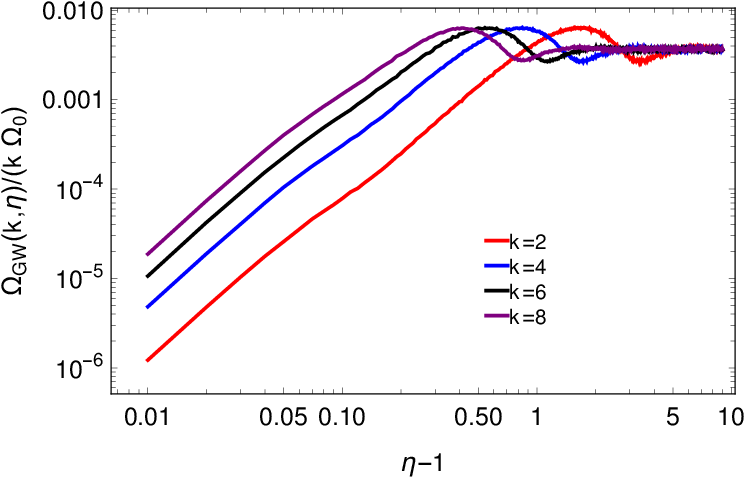}
\includegraphics[scale=0.6]{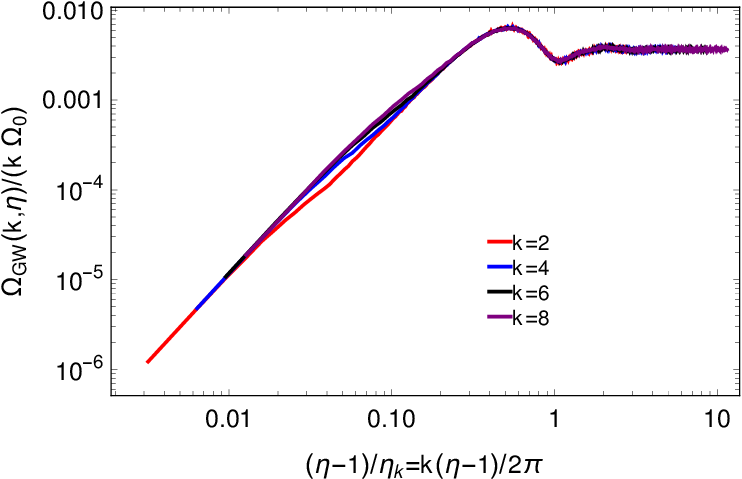}
\end{center}\caption[]{In this figure, we show the evolution of $\Omega_{GW}/(k\Omega_0)$ with time for $k=2,4,6,$ and 8. In the right panel, we normalized the time with $\eta_k=2\pi/k$.
}\label{time_evolution_of_spectrum_wo_exp_kp30}\end{figure*}

In this appendix, we present the time evolution of $\OmGW/(k\Omega_0)$
for wave numbers within the linear scaling region of the GW spectrum.
Specifically, in \Fig{time_evolution_of_spectrum_wo_exp_kp30}, we show
the time-dependent behavior of $\OmGW/(k\Omega_0)$ for values of
$k$ equal to 2, 4, 6, and 8, for the case in which $\kp=30$.
In the right-hand panel of this figure, we normalize the abscissa using
a wave number-dependent time scale denoted as $\eta_k=2\pi/k$.
This normalization suggests an inverse relationship between the saturation
time scale and wave number.

\section{Effect of kinetic energy growth period on the GW spectrum}
\label{TimeEvolutionKineticEnergy}

In \Sec{result1}, we assumed that the
kinetic energy remains unchanged throughout the evolution.
We have found that such a case produces a GW spectrum
characterized by a steep decline just below its peak, smoothly
transitioning to linear scaling at lower wave numbers.
However, in the context of a phase transition, the kinetic energy
increases and reaches a constant over a certain time span.

\begin{figure*}\begin{center}
\includegraphics[scale=1.22]{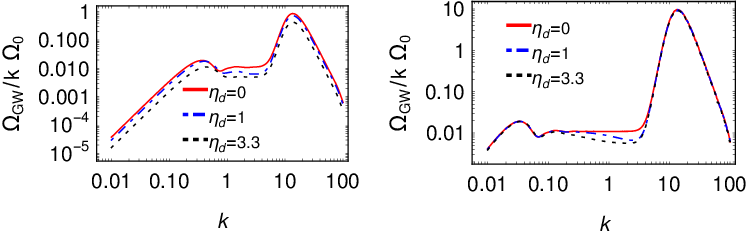}
\includegraphics[scale=1.22]{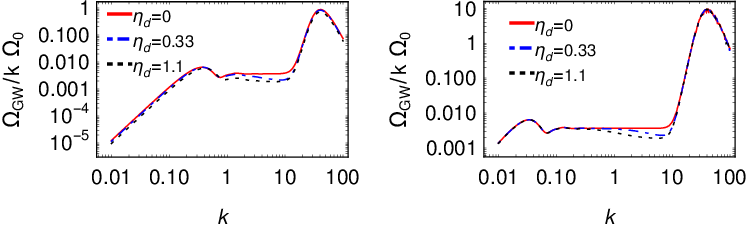}
\end{center}\caption[]{
Compensated GW spectra, $\OmGW/k\Omega_0$, for non-expanding cases with different values of growth time $\etad$, peak wave number 
$\kp$, evaluated at conformal time $\etae$ for $\etae=10$ and $100$ in the left- and right-hand panels.
In the top (bottom) panels, we have $\kp=10$ ($\kp=30$).
}\label{kin_spec_time_evolution_kp10_new}\end{figure*}

In this appendix, we consider a simplified model for the time evolution
of the kinetic energy and study its effect on the linearly rising part
of the GW spectrum at low wave numbers.
We assume that the kinetic energy evolves such that
the quantity $\Delta$ has the form 
\begin{align}
\Delta(\eta,\etai,\etad,k,\omega,\tilde{\omega})=&\frac{1}{2}\int_{\etai}^{\etae} d\eta_1 \int_{\etai}^{\etae}d\eta_2 \cos(k(\eta_1-\eta_2)) \cos(\omega(\eta_1-\eta_2))  \cos(\tilde{\omega}(\eta_1-\eta_2))
\nonumber\\
&\times\left[1-e^{-(\eta_1-\etai)/\etad}\right]\left[1-e^{-(\eta_2-\etai)/\etad}\right]\\
&=\frac{1}{8}\sum_{\pm\pm} \left[g_c(\eta, 
\etai, \etad, \omega_{\pm\pm})^2+g_s(\eta, 
\etai, \etad, \omega_{\pm\pm})^2\right],
\end{align}
where 
\begin{eqnarray}
  g_c(\eta,\etai,\etad,\omega)&=& \frac{\etad \left(\sin\omega\etai+\omega\etad\cos\omega\etai\right)-\etad e^{-(\eta-\etai)/\etad}\left(\sin\omega\eta +\omega\etad\cos\omega\eta\right)}{1+\left(\omega\etad\right)^2}\nonumber\\ 
  &&+\frac{\cos\omega\eta -\cos\omega\etai}{\omega},\\
  g_s(\eta,\etai,\etad,\omega)&=&-\frac{\etad \left(\cos\omega\etai-\omega\etad\sin\omega\etai\right)+\etad e^{-(\eta-\etai)/\etad}\left(-\cos\omega\eta+\omega\etad\sin\omega\eta \right)}{1+\left(\omega\etad\right){}^2}\nonumber\\
  &&+\frac{\sin\omega\eta-\sin\omega\etai}{\omega},
\end{eqnarray}
and $\etad$ is a parameter that controls the growth rate of the kinetic energy.
One can check that in the limit of instantaneous appearance, $\etad \to 0$, one recovers the result
\begin{equation}
    g_c(\eta, 
\etai, \omega)^2+g_s(\eta, 
\etai, \omega)^2 = \frac{4\sin^2[\omega(\eta-\etai)/2]}{\omega^2} ,
\end{equation}
consistent with \Eqs{FourTerms}{e:Sdef}.

In \Fig{kin_spec_time_evolution_kp10_new}, we show $\OmGW/k \Omega_0$ for different
values of $\kp$, $\etad$, and $\etae$. The upper panel shows the spectrum for $\kp=10$
and the lower panel corresponds to for $\kp=30$. Each panel is further divided: the
left side shows $\OmGW/k \Omega_0$ at $\etae=10$, whereas the right-hand side corresponds
to $\etae=100$. In each plot, the red, blue, and black curves correspond to the 
case with no time evolution ($\etad\to 0$),
$\etad=1/(0.1 \kp)$, and $\etad=1/(0.03\kp)$, respectively.
From this figure, it is evident that the time evolution affects the
amplitude of the shallow part of the GW spectrum towards the peak, 
thus slightly reducing the slope.
However, this effect reduces as we 
decrease the value of $\etad$.

\section{Case of an non-uniform initial energy density}\label{appendixd}
\begin{figure*}\begin{center}
\includegraphics[scale=1.3]{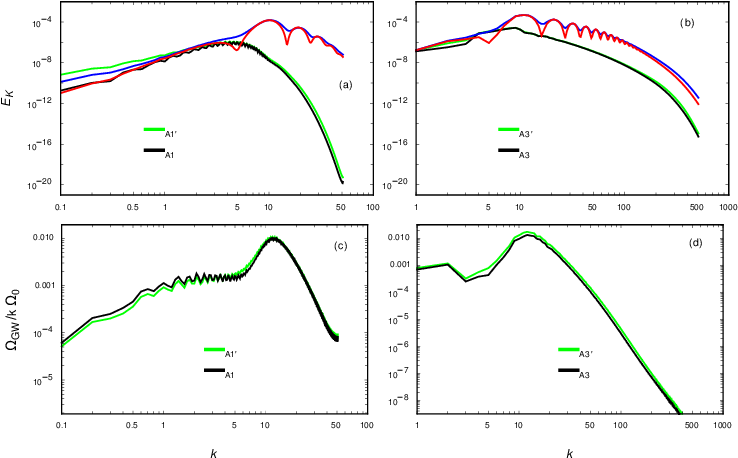}
\end{center}\caption[]{Kinetic and normalized GW spectra at different times. The red and blue curves represent the spectra at $t=1.5$, while the black and green curves correspond to $t=41.0$. }\label{acoustic_vs_acoustic}\end{figure*}
To consider the case of a non-uniform initial energy density such that
$\widetilde{\ln\rho}(k) = \kk \cdot \tilde{\uu}/(c_s^2 k)$, we have conducted
new simulations A1$'$ and A3$'$, analogous to the runs A1 and A3.
In \Fig{acoustic_vs_acoustic}, we provide a comparison between these
simulations.
The red and black curves represent the uniform case, while the red and
green curves correspond to the non-uniform case at simulation times
$t=1.5$ and $t=41$, respectively.
From this figure, it is evident that the ringing effect is reduced in
the non-uniform case.
Additionally, it is worth noting that the resulting GW spectra at $t=41$
exhibit negligible differences between the two cases.
Consequently, we conclude that initializing the energy density in
accordance with the continuity equation is important, although its
impact on the resulting GW spectrum is negligible.

\bibliographystyle{JCAP}
\bibliography{ref}
\end{document}